\documentstyle[12pt]{article}
\begin{document}

\newcommand{\be}{\begin{equation}}
\newcommand{\ee}{\end{equation}}

{}~~~~
\vspace{1cm}

\centerline{\bf         EXACT EQUATIONS}
\vspace{3mm}

\centerline{\bf         FOR VACUUM CORRELATORS IN FIELD THEORY}
\vspace{10mm}

\centerline{            D.V.ANTONOV}

\centerline{\it Moscow Engineering Physics Institute,}
\centerline{\it Moscow, Kashirskoe highway 31, 115409, Russia}
\vspace{3mm}

\centerline{            YU.A.SIMONOV}

\centerline{\it Institute of Theoretical and Experimental Physics,}
\centerline{\it Moscow, ~B.Cheremushkinskaya 25, ~117259, ~Russia}
\vspace{10mm}

\begin{abstract}

Stochastic quantization is used to derive exact equations, connecting
multilocal field correlators in the $\varphi^3$ theory and gluodynamics.
Perturbative expansion of the obtained equations in the lowest orders is
presented.
\end{abstract}


\section{Introduction}

Exact determination of nonperturbative ef\/fects is one of the main problems
in  f\/ield theory nowadays. We suggest to look for them in exact
equations (eqs.) for vacuum correlators, which we derive, using
stochastic quantization method [1].

In QCD the nonperturbative correlators are an essential part of
operator product expansion (OPE), QCD sum rules [2] and, more
recently, of the method of  vacuum correlators (MVC) [3]. In the
latter correlators are used as a dynamical input, and one needs to
obtain eqs. for them, starting from the lagrangian.

There exist in literature two types of exact eqs. First is Dyson--Schwinger
eqs.[4], which, however, are not applicable to QCD  in  the
confining phase, because of the lack of gauge invariance. The other
one is Makeenko-Migdal eqs.[5], which are closer to our eqs., but
written not for correlators, but for Wilson loops in the large $N_c$
limit. We hope, that the eqs., found in this letter for correlators,
can be used more directly for the purposes, stated above.

Sections 2 and 3 are devoted to the $\varphi^3$  theory, and in
Section 4 an alternative approach is applied to gluodynamics.


\section{$\varphi^3$ theory}


{}From the Euclidean action
$$
S=\int
dx\left[\frac{1}{2}\Bigl(\partial_{\mu}\varphi\Bigr)^2+
\frac{m^2}{2}\varphi^2-\frac{g}{3}\varphi^3\right]
$$
 in the stochastic
 quantization method one obtains the Langevin equation
\be
\dot{\varphi}(x,t)+\Bigl(m^2-\partial^2-g\varphi(x,t)\Bigr)
\varphi(x,t) = \eta(x,t) ~,
\ee
where $<\eta(x,t)\eta(x',t')>~= 2\delta(x-x')\delta(t-t')$. In the
absence of constant classical solutions the retarded Green function
of (1) is
$$
G(x,y,t) ~=~ \theta(t)<x|e^{-(m^2-\partial^2
-g\varphi)t}|y>~ =
$$
$$
=~ \theta(t)\int
(Dz)_{xy}
exp\left(-m^2t-\int\limits^t_0\frac{\dot{z}^2}{4}d\xi+g
\int\limits^t_0\varphi(z(\xi),\xi)d\xi \right) ~,
$$
where
$$
(Dz)_{xy}=\lim_{N\to\infty}\prod^N_{n=1}\frac{d^4z(n)}
{(4\pi\varepsilon)^2},~~~~
 N\varepsilon=t,~~~~ z(\xi=0)=y,~~~~ z(\xi=t)=x ~.
$$
Here we used the path integral  Feynman--Schwinger representation.
Thus, the solution of (1) has the form
$$
\varphi
(x,t)=\int\limits^t_0dt'\int dy (Dz)_{xy} K_z(t,t')F_z(t,t')
\eta(y,t') ~,
$$
where
$$
K_z(t,t')=\theta(t-t')
exp\left(-m^2(t-t')-\int\limits^t_{t'}\frac{\dot{z}^2(\xi)}
{4}d\xi \right) ~,
$$
$$
  F_z(t,t')=exp~ \left(g\int\limits^t_{t'} \varphi
  (z(\xi),\xi)d\xi \right) ~.
$$

To obtain the system of eqs. for
$<\varphi>,~ <\varphi\eta>,~ <\varphi\varphi>,~ <\varphi\varphi
\eta>, ...~$ we use the generating functional
$$
  Z[J]=<F_z(t,t')F_{\bar{z}}(\bar{t},\bar{t}')exp\left(\int du dt
  J(u,t)\eta(u,t)\right) >
$$
and apply to it cumulant expansion [6]:
$$
Z[J]=exp\sum^{\infty}_{n=1}\frac{1}{n!}
\ll \left(g\int\limits^t_{t'}\varphi(z(\xi),\xi)d\xi+
g\int\limits^{\bar{t}}_{\bar{t}'}\varphi \Bigl(
\bar{z}(\bar{\xi}),\bar{\xi}\Bigr)d\bar{\xi}+
\int dudtJ(u,t)\eta(u,t) \right)^n\gg ~.
$$
Then
$$
<F_z(t,t')\eta(y,t')>=\frac{\delta Z[J]}{\delta
J(y,t')}\left|_{\stackrel{J=0}
{F_{\bar{z}}(\bar{t},\bar{t}')=1}} \right. =
$$
$$
=\left\{\sum^{\infty}_{n=1}\frac{1}{(n-1)!}\ll
\left(g\int\limits^t_{t'}\varphi(z(\xi),\xi)d\xi\right)^{n-1}
\eta(y,t')\gg \right\} < F_z(t,t') > ~,
$$
$$
<F_z(t,t')\eta(y,t')\eta(u,t)>=\frac{\delta^2 Z[J]}{\delta
J(y,t')\delta J(u,t)}\left|_{\stackrel{J=0}
{F_{\bar{z}}(\bar{t},\bar{t}')=1}} \right. =
$$
$$
=\left\{\sum^{\infty}_{n=2}\frac{1}{(n-2)!}\ll
\left(g\int\limits^t_{t'}\varphi(z(\xi),\xi)
 d\xi\right)^{n-2}\eta(y,t')\eta(u,t)\gg + \right.
$$
$$
+ \left[ \sum^{\infty}_{n=1}\frac{1}{(n-1)!} \ll
\left(g\int\limits^t_{t'}\varphi(z(\xi),\xi)d\xi\right)^{n-1}\eta(y,t')\gg
\right] \cdot
$$
$$
\cdot  \left. \left[ \sum^{\infty}_{k=1}\frac{1}{(k-1)!}
\ll \left(g\int\limits^t_{t'}\varphi(z(\xi),\xi)d\xi\right)^{k-1}
\eta(u,t)\gg\right] \right\} < F_z(t,t') >
$$
and so on. In the bilocal approximation (i.e. omitting all the
cumulants higher than quadratic), one gets:
$$
<\varphi(x,t)> ~=
$$
\be
= g\int\limits^t_{0}dt'\int dy
 (Dz)_{xy}K_z(t,t')\int\limits^t_{t'}d\xi<\varphi(z(\xi),\xi)\eta(y,t')>
 <F_z(t,t')> ~,
\ee
$$
<\varphi(x,t)\eta(\bar{x},\bar{t})> ~=
$$
$$
=2\int(Dz)_{x\bar{x}}K_z(t,\bar{t})<F_z(t,\bar{t})>+
g^2\int\limits^t_{0}dt'\int dy
 (Dz)_{xy}K_z(t,t') \cdot
$$
\be
\cdot \int\limits^t_{t'}d\xi
 <\varphi(z(\xi),\xi)\eta(\bar{x},\bar{t})>\int\limits^t_{t'}d\xi'<
 \varphi(z(\xi'),\xi')\eta(\bar{x},\bar{t})><
 F_z(t,t')> ~,
\ee
$$
<\varphi(x,t)\varphi(\bar{x},\bar{t})> ~=
$$
$$
=g^2\int\limits^t_0dt'\int\limits^{\bar{t}}_0d\bar{t}'
\int
dyd\bar{y}(Dz)_{xy}
(D\bar{z})_{\bar{x}\bar{y}}K_z(t,{t}')K_{\bar{z}}(\bar{t},\bar{t}') \cdot
$$
$$
\cdot \left[\int\limits^t_{t'}d\xi
 <\varphi(z(\xi),\xi)\eta(y,t')> +
 \int\limits^{\bar{t}}_{\bar{t'}}d\xi'<
 \varphi(\bar{z}(\xi'),\xi')\eta(y,t')> \right] \cdot
$$
$$
\cdot  \left[ \int\limits^t_{t'} d \xi < \varphi (z (\xi), \xi) \eta
  (\bar y, \bar t') > +
  \int\limits^{\bar{t}}_{\bar{t'}}d\xi'<
 \varphi(\bar{z}(\xi'),\xi')\eta(\bar{y},\bar{t'})>\right] \cdot
$$
\be
\cdot <F_z(t,t')F_{\bar{z}} ( \bar{t},\bar{t'}) > + 2
\int\limits^t_0dt'\int dy (Dz)_{xy}(D\bar{z})_{\bar{x}y}
K_z(t,t')K_{\bar{z}}(\bar{t},t')<F_z(t,t')F_{\bar{z}}(\bar{t},t')>,
\ee
where
$$
<F_z(t,t') > ~=~
= exp~ \left(g\int\limits^t_{t'}d\xi
<\varphi(z(\xi),\xi)> + \right.
$$
\be
+ \frac{g^2}{2}\int\limits^t_{t'}d\xi
\int\limits^t_{t'}d\xi'\left(<\varphi(z(\xi),\xi)
\varphi(z(\xi'),\xi')> -  <\varphi(z(\xi),\xi)>
<\varphi(z(\xi'),\xi')> \right) \left. \right) ~,
\ee
$$
<F_z(t,t')F_{\bar{z}}(\bar{t},\bar{t'})> ~=~
{}~exp \left\{ g\int\limits^t_{t'}d\xi
<\varphi(z(\xi),\xi)>+ \right.
$$
$$
 + g\int\limits^{\bar{t}}_{\bar{t'}}d\bar{\xi}
<\varphi(\bar{z}(\bar{\xi}),\bar{\xi})>+
\frac{g^2}{2}
\left[ \int\limits^t_{t'}d\xi
\int\limits^t_{t'}d\xi'<
\varphi(z(\xi),\xi)
\varphi(z(\xi'),\xi')> + \right.
$$
$$
 + \int\limits^{\bar{t}}_{\bar{t'}}d\bar{\xi}
\int\limits^{\bar{t}}_{\bar{t'}}d\bar{\xi'}
<\varphi(\bar{z}(\bar{\xi}),\bar{\xi})
\varphi(\bar{z}(\bar{\xi'}),\bar{\xi'})>+
 2\int\limits^t_{t'}d\xi\int\limits^{\bar{t}}_{\bar{t'}}d{\bar{\xi}}
<\varphi({z}({\xi}),{\xi})
\varphi(\bar{z}(\bar{\xi}),\bar{\xi})> -
$$
\be
\left. \left. - \left( \int\limits^{t}_{t'}d{\xi}
<\varphi({z}({\xi}),{\xi})> +
\int\limits^{\bar{t}}_{\bar{t'}}d\bar{\xi}
<\varphi(\bar{z}(\bar{\xi}),\bar{\xi})>\right)^2 \right] \right\} ~.
\ee
Eqs.(2)-(6) is the minimal closed set of eqs. for $<\varphi>,
{}~<\varphi\eta>, ~<\varphi\varphi>$.

Note, that $<\varphi(x)\varphi(x')>_{vac}=\lim_{{\stackrel{t\to\infty}
{\stackrel{t'\to\infty}{(t-t')fixed}}}} <\varphi(x,t)\varphi
(x',t')>_\eta.$ It means, that without only suppositions about the
structure of the real va\-cu\-um, we obtained eqs. for physical
correlators.


\section{Perturbative expansion of eqs.(2)-(6) }


Expanding the right hand size of eqs.(2)-(6) in powers of
$g$, we get in the lowest order
$$
<\varphi(x,t)>^{(0)}=0 ~,
$$
$$
<\varphi(x,t){\eta}(\bar{x},\bar{t})>^{(0)}=
2\int(Dz)_{x\bar{x}}K_z(t,\bar{t})=
-\frac{1}{8\pi^2(\bar{t}-t)^2}e^{\frac{(\bar{x}-x)^2}{4(\bar{t}-t)}+m^2
(\bar{t}-t)} ~.
$$
The last term on the right hand size of (4) can be written as
$$
\int dy<x|KF|y><y|KF|\bar{x}> ~=~ <x|KFKF|\bar{x}> ~,
$$
 and for $t=\bar{t}$ one
obtains for  this term
$$
\int\limits^{2t}_{0}dt_1(Dz)_{x\bar{x}}
K_z(t_1,0)<\bar{F}_z(t_1,0)> ~,
$$
where
\be
\bar{F}_z(t_1,0)=exp\left[\int\limits^{\frac{t_1}{2}}_0\varphi\left(
z(\xi),\xi+t-\frac{t_1}{2}\right)
d\xi+\int\limits^{t_1}_{\frac{t_1}{2}}\varphi\Bigl(z(\xi),\xi+
t-t_1\Bigr)d\xi \right] ~.
\ee

Note, that in the asymptotical regime, when one drops out the dependence on
$\xi$ in $\varphi(z(\xi),\xi)$, we have $\bar{F}_z\to F_z$. Using
(7),
$$
<\varphi(x,t)\varphi(\bar{x},t)>^{(0)}=
$$
$$
 = \int\limits^{2t}_0
dt_1\int(Dz)_{x\bar{x}}K_z(t_1,0)=\int\frac{dp}{(2\pi)^4}
\frac{e^{ip(x-\bar{x})}(1-e^{-2(p^2+m^2)t})}{p^2+m^2} ~.
$$
This expression tends to the propagator
of a free boson in the limit $t\to\infty$.

In the next order we have
$$
<\varphi(x,t)>^{(1)} ~=~2g\int\limits^t_0dt'\int dy(Dz)_{xy}K_z(t,t')
 \int\limits^t_{t'}d\xi\int(D\bar{z})_{z(\xi)y}K_{\bar{z}}(\xi,t') =
$$
$$
=2g\int\limits^t_0dt'\int dy(Dz)_{xz(\xi)}K_z(t,\xi)(Dz)_{z(\xi)y}
K_z(\xi,t')dz(\xi)\int\limits^t_{t'}d\xi
\int\frac{dp}{(2\pi)^4}
e^{ip(z(\xi)-y)-(p^2+m^2)(\xi-t')} =
$$
$$
=g\int\frac{dp}{(2\pi)^4}\frac{1}{m^2(p^2+m^2)}+
{\rm {\underline{\underline{0}}}}
(e^{-m^2t}) ~,
$$
 that in the limit $t\to \infty$ corresponds to the lowest
order tadpole diagram.

Keeping $ <\varphi>^{(1)}$ in the last term on the right hand size of (4), we
obtain
$$
 <\varphi(x,t)\varphi(\bar{x},t)>^{(1)}= \int\limits^{2t}_0
dt_1\int\frac{dp}{(2\pi)^4}
e^{ip(x-\bar{x})-\Bigl(p^2+m^2-g<\varphi>^{(1)}\Bigr)t_1}=
$$
\be
=\int\frac{dp}{(2\pi)^4}\frac{e^{ip(x-\bar{x})}}{p^2+m^2-g<\varphi>^{(1)}}+
{\rm {\underline{\underline{0}}}}
(e^{2(g<\varphi>^{(1)}-m^2)t}) ~.
\ee
Expanding (8) in powers of $g$, one obtains an inf\/inite set of
diagrams.

Note, that the unstability of vacuum in the $\varphi^3$ theory leads to
divergence of (8) if $g<\varphi>^{(1)}>m^2$, so  that the stochastic
process has not limiting  equilibrium.


\section{Gluodynamics}


We shall start from the Langevin equation
\be
\dot{A}_{\mu}^a=(D_{\lambda}F_{\lambda\mu})^a-\eta^a_{\mu},
\ee
where
$F^a_{\lambda\mu}=\partial_{\lambda}A^a_{\mu}-\partial_{\mu}A^a_{\lambda}+
gf^{abc}A_{\lambda}^bA^c_{\mu},(D_{\lambda}F_{\lambda\mu})^a=\partial_{\lambda}
F^a_{\lambda\mu}+gf^{abc}A^b_{\lambda}F^c_{\lambda\mu}$, and the
sign of $\eta^a_{\mu}$ is changed.  Let's use Schwinger gauge [7]
$A_{\mu}(x,t)(x-x_0)_{\mu}=0$, in which
$A_{\mu}(x,t)=\int^x_{x_0}dz_{\nu}\alpha(z,x)F_{\nu\mu}(z,t)$, where
$\alpha(z,x)=\frac{(z-x_0)_{\lambda}(x-x_0)_{\lambda}}{(x-x_0)^2},~x_0$
is an arbitrary point (here and later in all the integrals of the
type $\int^x_{x_0}dz_{\nu}$ the path of integration is a straight
line), and introduce the generating functional
$$\Phi_{\beta}(t)=P
{}~exp~ig\oint_{c}dx_{\mu}(\int^x_{x_0}dz_{\nu}\alpha(z,x)F_{\nu\mu}
(z,t)+\beta n_{\mu}
(x,t))$$, where
$n_{\mu}(x,t)=\int^t_0\eta_{\mu}(x,t')dt',C$ is some fixed closed
contour.

According to (9),
\be
tr\frac{\partial}{\partial t}<\Phi_{\beta}(t)>=ig tr \oint_{c}
du_{\mu}<\Phi_{\beta}(t)V_{\mu}(\beta,u,x_0,t)>,
\ee
where
$$V_{\mu}(\beta,u,x_0,t)=
\Phi(x_0,u,t)(D_{\lambda}
F_{\lambda\mu}(u,t)+(\beta-1)\eta_{\mu}(u,t))\Phi(u,x_0,t),
 $$
 $$\Phi(u,x_0,t )=P~ exp~ig \int^u_{x_0}A_{\mu}(z,t)dz_{\mu}.$$

 Applying to both sides of (10) cumulant expansion, using the formula
 [6] $$<e^AB>=<e^A>(<B>+\sum^{\infty}_{n=1}\frac{1}{n!}\ll
 A^nB\gg),$$ where $A$ and $B$ are two statistically dependent
 matrixes, and putting $\beta=1$, one obtains in the bilocal
 approximation $$
 tr(\int^y_{x_0}dz_{\lambda}\alpha(z,y)\int^u_{x_0}dx_{\rho}\alpha(x,u)
 \frac{\partial}{\partial t}
 <F_{\lambda\nu}(z,x_0,t)F_{\rho\mu}(x,x_0,t)>+\int^y_{x_0}dz_{\lambda}\alpha
 (z,y)\int^t_0 dt'\frac{\partial}{\partial t}\cdot
 $$
 $$
 <F_{\lambda\nu}(z,x_0,t)\eta_{\mu}(u,x_0,t,t')>+
 \int^u_{x_0}dx_{\rho}\alpha
 (x,u)\int^t_0 dt'\frac{\partial}{\partial t}
 <F_{\rho\mu}(x,x_0,t)\eta_{\nu}(y,x_0,t,t')>+
 $$
 $$+\int^y_{x_0}dz_{\lambda}\alpha
 (z,y)
 <F_{\lambda\nu}(z,x_0,t)\eta_{\mu}(u,x_0,t,t)>+ $$
 $$+\int^u_{x_0}dx_{\rho}\alpha
 (x,u)
 <F_{\rho\mu}(x,x_0,t)\eta_{\nu}(y,x_0,t,t)>+
 $$
 $$
 +\int^t_0dt'(<\eta_{\nu}(y,x_0,t,t)\eta_{\mu}(u,x_0,t,t')>+
 <\eta_{\nu}(y,x_0,t,t')\eta_{\mu}(u,x_0,t,t)>)+
 \int^t_0dt'\int^t_0dt^{''}\cdot
  $$
 $$
 \cdot \frac{\partial}{\partial t}
 <\eta_{\nu}(y,x_0,t,t')\eta_{\mu}(u,x_0,t,t^{''})>)=
 2tr(\int^y_{x_0}dz_{\lambda}\alpha(z,y)\frac{\partial}{\partial
 u_{\rho}}<F_{\lambda\nu}(z,x_0,t)F_{\rho\mu}(u,x_0,t)>+
 $$
 \be
 +\int^t_0dt^{'}\frac{\partial}{\partial
 u_{\rho}}<F_{\rho\mu}(u,x_0,t)
 \eta_{\nu}(y,x_0,t,t')>),
 \ee
 where
 $$
 F_{\lambda
 \nu}(z,x_0,t)=\Phi(x_0,z,t)F_{\lambda\nu}(z,t)
 \Phi(z,x_0,t),\eta_{\mu} (u,x_0,t,t')=\Phi(x_0,u,t)\cdot
 $$
 $$\cdot \eta_{\mu}(u,t')\Phi(u,x_0,t).$$

 Differentiating (10) by $\beta$, we get
 \be
 tr\frac{\partial}{\partial
 t}<\Phi_{\beta}(t)n_{\mu}(u,x_0,t)>=tr(ig\oint_{c}dz_{\nu}
 <\Phi_{\beta}(t)n_{\nu}(z,x_0,t)
 V_{\mu}(\beta,u,x_0,t)>+
 \ee
 $$+<\Phi_{\beta}(t)\eta_{\mu}(u,x_0,t)>),
 $$
 that in bilocal approximation yields
 $$
 tr\frac{\partial}{\partial t}
 (\int^y_{x_0}dz_{\lambda}\alpha(z,y)
 <F_{\lambda\nu}(z,x_0,t)n_{\mu}(u,x_0,t)>+
 <n_{\nu}(y,x_0,t)n_{\mu}(u,x_0,t)>)=
 $$
 $$
 =tr(<\Phi(x_0,u,t)(D_{\lambda}F_{\lambda\mu}(u,t))\Phi(u,x_0,t)
 n_{\nu}(y,x_0,t)>+
 \int^y_{x_0}dz_{\lambda}\alpha(z,y)
 <F_{\lambda\nu}(z,x_0,t)\cdot
 $$
 $$
 \cdot \eta_{\mu}(u,x_0,t)>+
 <n_{\nu}(y,x_0,t)\eta_{\mu}(u,x_0,t)>).$$
 Noticing that
 $$
 tr(<\Phi(x_0,u,t)(D_{\lambda}F_{\lambda\mu}(u,t))\Phi(u,x_0,t)
 n_{\nu}(y,x_0,t)>= $$
 $$
 =\frac{\partial}{\partial u_{\lambda}} tr
 <F_{\lambda\mu}(u,x_0,t)n_{\nu}(y,x_0,t)>+
 ig \int^u_{x_0}dx_{\rho}\cdot
 $$
 $$
 \cdot \alpha(x,u)
 (tr
 <F_{\lambda\mu}(u,x_0,t)
 n_{\nu}(y,x_0,t)
 F_{\lambda\rho}(x,x_0,t)>-
 tr <F_{\lambda\mu}(u,x_0,t) F_{\lambda\rho}(x,x_0,t)\cdot
 $$
 $\cdot n_{\nu}(y,x_0,t)>)
 $ and using the expression for threelocal path--ordered cumulant
 $\ll 123\gg=<123>-<1><23>-<12><3>+<1><2><3>$ [7], one obtains
 $$
 tr(\int^y_{x_0}dz_{\lambda}\alpha(z,y)
 \frac{\partial}{\partial t}
 <F_{\lambda\nu}(z,x_0,t)\eta_{\mu}(u,x_0,t,t')>+
 \int^t_{0}dt^{''}
 \frac{\partial}{\partial t}
 <\eta_{\nu}(y,x_0,t')\eta_{\mu}(u,x_0,t,t^{''})>-
 $$
 \be
  -\frac{\partial}{\partial u_{\lambda}}
 <F_{\lambda\mu}(u,x_0,t)\eta_{\nu}(y,x_0,t,t^{'})>)=-
 tr <\eta_{\nu}(y,x_0,t,t)\eta_{\mu}(u,x_0,t,t')>.
 \ee

 Differentiating (12) by $\beta$, putting
 $\beta=1$ and using the expression for
 fourlocal path--ordered cumulant
 $\ll 1234\gg=<1234>-\ll 123\gg$ $<4>-<1>\ll
 234\gg-\ll 12\gg\ll 34\gg-\ll
 12\gg<3><4>-<1>\ll 23\gg<4>-<1><2>\ll
 34\gg-<1><2><3><4>$, we have in bilocal
 approximation:
 $$
 tr(\int^t_0dt^{''}\frac{\partial}{\partial
 t}<\eta_{\nu}(y,x_0,t,t')\eta_{\mu}(u,x_0,t,t^{''})>
 +<\eta_{\nu}(y,x_0,t,t)\eta_{\mu}
 (u,x_0,t,t')>-
 $$
 $$-<\eta_{\nu}(y,x_0,t,t')\eta_{\mu}(u,x_0,t,t)>)=
 g^2tr(\int^u_{x_0}
 dx_{\sigma}\alpha(x,u)<F_{\rho\mu}(u,x_0,t)F_{\rho\sigma}(x,x_0,t)>\cdot
 $$
 $$\cdot \oint_{c}
 dz_{\lambda}\int^t_0dt^{''}<\eta_{\lambda}(z,x_0,t,t')\eta_{\nu}
 (y,x_0,t,t^{''})>-
  \oint_{c}
 dz_{\lambda}\int^t_0dt^{''}<F_{\rho\mu}(u,x_0,t)\eta_{\lambda}
 (z,x_0,t,t')>\cdot
 $$
 \be
 \cdot \int^u_{x_0}
 dx_{\sigma}\alpha(x,u)
 <\eta_{\nu} (y,x_0,t,t^{''})
 F_{\rho\sigma}(x,x_0,t)>)
 \ee

 Eqs. (11),(13) and (14) is the  minimal closed set of eqs. for
 correlators
  $$<F_{\lambda\nu}(z,x_0,t)F_{\rho\mu}(x,x_0,t)>,
 <F_{\lambda\nu}(z,x_0,t)\eta_{\mu}(u,x_0,t,t')>,$$
 $$<\eta_{\nu}(y,x_0,t,t')\eta_{\mu}(u,x_0,t,t^{''})>.$$

 Note, that beyond obtained eqs. (and all other eqs., one is able to
 get, using $\Phi_{\beta}$) there are some additional relations,
 connecting gauge--invariant correlators. Let
 $G_{\mu_1...\mu_n}(x_1,t_1,...,x_n,t_n,x_0,t)$ be a product of some
 number of $F_{\mu\nu}$ and (or) $\eta_{\mu}$, which are given in the
 points $x_1,...,x_n$ at the moments $t_1,...,t_n$ of fictitious time
 respectively, and all the parallel transporters between $x_0$ and
 each of these points are given at the same moment $t$. Then, using
 nonabelian Bianchi identities $D_{\lambda}\tilde{F}_{\lambda\mu}=0$,
 we have [7]:
 $$
 tr\frac{\partial}{\partial
 x_{\lambda}}<\tilde{F}_{\lambda\mu}(x,x_0,t')G_{\mu_1...\mu_n}>=
 ig tr\int^x_{x_0} dz_{\rho}\alpha (z,x)(<
 \tilde{F}_{\lambda\mu}(x,x_0,t')
 {F}_{\lambda\rho}(z,x_0,t)
 G_{\mu_1...\mu_n}>-
 $$
 $$
 -<\tilde{F}_{\lambda\mu}(x,x_0,t')
 G_{\mu_1...\mu_n}{F}_{\lambda\rho}(z,x_0,t) >),~~{\rm{where}}~~x\neq
 x_1,...,x\neq x_n.  $$

 To check ourselfs, let's consider eqs. (11), (13) and (14) in the
 order $g^0$. Looking for $<A_\nu^a(y,t)\eta^b_{\mu}(u,t')>$ in the
 form $\delta^{ab}d_{\mu\nu}(z,\tau),$ where\\
 $z
 =u-y,\tau=|t-t'|,d_{\nu\mu}(z,\tau)=d_{\mu\nu}(z,\tau),d_{\mu\nu}(-z,\tau)=
 d_{\mu\nu}(z,\tau)$,  one obtains from (13):
 $(\delta_{\mu\lambda}\frac{\partial}{\partial\tau}+
 \frac{\partial^2}{\partial z_{\mu}\partial z_{\lambda}}-
 \delta_{\mu\lambda}\partial^2)d_{\lambda\nu}(z,\tau)=
 -2\delta_{\mu\nu}\delta(z)\delta(\tau)$ and, thus, \\
  $d_{\mu\nu}(k,\tau)=
  -2\theta(\tau)(T_{\mu\nu}e^{-k^2\tau}+L_{\mu\nu})$, where
 $T_{\mu\nu}=\delta_{\mu\nu}-\frac{k_{\mu}k_{\nu}}{k^2},
 L_{\mu\nu}=\frac{k_{\mu}k_{\nu}}{k^2}.$

  Looking for $<A_{\nu}^a(y,t)A^b_{\mu}(u,t)>$ in the
 form $\delta^{ab}h_{\mu\nu}(z,t),$ where \\
 $h_{\nu\mu}(z,t)=h_{\mu\nu}(z,t),
 h_{\mu\nu}(-z,t)=h_{\mu\nu}(z,t)$,  we have  from (11): \\
 $(\frac{1}{2}\delta_{\mu\lambda}\frac{\partial}{\partial t}+
 \frac{\partial^2}{\partial z_{\mu}\partial z_{\lambda}}-
 \delta_{\mu\lambda}\partial^2)h_{\lambda\nu}(z,t)=
 -d_{\mu\nu}(z,0).$\\ So,
  $h_{\mu\nu}(k,t)=\frac{1}{k^2}
  T_{\mu\nu}(1-e^{-2k^2t})+2tL_{\mu\nu}$,
   that is the ordinary photon pro\-pa\-ga\-tor
   in the stochastic quantization method [1].

   \section{Conclusion}

   The main result of the present letter is eqs. (11), (13) and (14).
   They are explicitly gauge invariant and produce correct
   perturbative results in the lowest order.

    The principle of separation of perturbative and nonperturbative
    contributions in the obtained eqs., introduction of quarks and
    the problem of re\-gu\-la\-ri\-zation will be treated in the next
    publications.

    One of us (Yu.S.) would like to thank Yu.M.Makeenko for useful
    discussions.

    The work is supported by the Russian Fundamental Research
    Foundation, Grant No. 93-02-14937.

      \end{document}